\documentclass[12pt, aps,amsmath,onecolumn,groupedaddress,superscriptaddress,notitlepage,nofootinbib,prd]{revtex4-2}

\usepackage[utf8]{inputenc}
\usepackage{indentfirst}
\usepackage{amssymb}
\usepackage{amsmath,latexsym}
\usepackage{graphicx}
\usepackage{float}
\usepackage[margin=1.0in]{geometry}
\usepackage[mathscr]{euscript}
\usepackage{color}
\usepackage[dvipsnames]{xcolor}
\usepackage{soul}
\usepackage{slashed}
\usepackage{feynmf}
\usepackage{subcaption}
\usepackage{braket}
\usepackage[normalem]{ulem}
\usepackage{comment}
\usepackage{makecell}

\usepackage{graphics,setspace,epsfig,color}

\usepackage[vcentermath]{}
\usepackage{epsf}
\usepackage{amscd}
\usepackage{amsmath}
\usepackage{amsfonts}
\usepackage[utf8]{inputenc}
\usepackage{microtype}
\usepackage{amssymb}
\usepackage{braket} 
\usepackage{bbm}
\usepackage{bm} 
\usepackage{mathtools}
\usepackage{slashed} 
\usepackage[dvipsnames]{xcolor} 
\usepackage[colorlinks=true,backref=false, linktocpage=true,
citecolor=blue,urlcolor=blue,linkcolor=blue,pdfpagemode=UseOutlines]{hyperref}

\usepackage{tcolorbox}

\definecolor{mybrown}{rgb}{0.6,0.4,0.2}
\definecolor{myblue}{rgb}{0.6,0.4,0.2}
\definecolor{myred}{rgb}{0.8,0.2,0.2}

\newcommand{\beq}{\begin{equation}}
\newcommand{\eeq}{\end{equation}}
\newcommand{\bea}{\begin{eqnarray}}
\newcommand{\eea}{\end{eqnarray}}

\newcommand{\eq}[1]{Eq.~\ref{#1}}
\newcommand{\fig}[1]{Fig.~\ref{#1}}
\newcommand{\tab}[1]{Table~\ref{#1}}

\usepackage{dsfont}

\long\def\beqs#1\eeqs{\beq\begin{split} #1 \end{split}\eeq}

\newcommand{\nup}{{N^\uparrow}}
\newcommand{\nd}{{N^\downarrow}}
\newcommand{\xup}{x^\uparrow}
\newcommand{\xd}{x^\downarrow}
\newcommand{\Xup}{X^\uparrow}
\newcommand{\Xd}{X^\downarrow}

\definecolor{MyRed}{RGB}{153,0,13}

\PassOptionsToPackage{hyphens}{url}
\usepackage{multirow}
\usepackage{booktabs}

\usepackage{tikz}

\preprint{}

\begin{document}
\title{Trapped Fermions  Through Kolmogorov-Arnold Wavefunctions}

\author{Paulo F. Bedaque}
\email{bedaque@umd.edu}
\affiliation{Department of Physics,
University of Maryland, College Park, MD 20742}

\author{Jacob Cigliano}
\email{cigliano@terpmail.umd.edu}
\affiliation{Department of Physics,
University of Maryland, College Park, MD 20742}

\author{Hersh Kumar}
\email{hekumar@umd.edu}
\affiliation{Department of Physics,
University of Maryland, College Park, MD 20742}

\author{Srijit Paul}
\email{spaul137@umd.edu}
\affiliation{Department of Physics,
University of Maryland, College Park, MD 20742}

\author{Suryansh Rajawat}
\email{suryansh@terpmail.umd.edu}
\affiliation{Department of Physics,
University of Maryland, College Park, MD 20742}

\preprint{}


\begin{abstract}
We investigate a variational Monte Carlo framework for trapped one-dimensional mixture of spin-$\tfrac{1}{2}$ fermions using Kolmogorov-Arnold networks (KANs) to construct universal neural-network wavefunction ans\"atze.
The method can, in principle, achieve arbitrary accuracy, limited only by the Monte Carlo sampling and 
 was checked against exact results at sub-percent precision. 
 For attractive interactions, it captures pairing effects, and in the impurity case it agrees with known results. We present a method of systematic transfer learning in the number of network parameters, allowing for efficient training for a target precision. We vastly increase the efficiency of the method by incorporating the short-distance behavior of the wavefunction into the ans\"atz without biasing the method.
\end{abstract}
\maketitle

\section{Introduction}
Recent developments in Machine Learning (ML)  provide new methods for addressing long-standing challenges in quantum many-body physics. A particularly active area of research connects neural networks with Variational quantum Monte Carlo (VMC), leveraging the functional flexibility of neural architectures to construct highly expressive ans\"atz for many-body wavefunctions. These neural quantum states can approximate arbitrary functions \cite{Cybenko1989} and their use 
opens the possibility of transforming VMC into an ``exact" method, namely, one for which arbitrarily accurate results can be found in well understood limits (in this case, the number of parameters of the network and the  Monte Carlo samples).

The connection is underpinned by a direct formal analogy between VMC and unsupervised machine learning: both frameworks involve the iterative minimization of a scalar objective through the adjustment of model parameters. In VMC, this objective is the energy of the quantum system, and the model is a neural network-based wavefunction ans\"atz. The stochastic evaluation of energies and gradients via Monte Carlo sampling further aligns with the optimization methods used in large-scale ML.
This analogy can be extended by observing a common practice in ML: the introduction of a regularization term to the cost function to discourage overfitting. Such a term, which penalizes highly oscillatory functions, can be mathematically identical to the kinetic energy operator in quantum mechanics. Through this lens, the problem of finding a quantum ground state is recast as the minimization of the potential energy, with a regularization term—whose strength is set by the value of $\hbar$--favoring smooth wavefunctions.

The parametrization of ground state wavefunctions using neural networks was initially demonstrated for spin systems with discrete variables, using models such as Restricted Boltzmann Machines \cite{carleo, saito_2017, choo_2018}. This work was later extended to continuous systems of particles in real space, with comprehensive studies of long-range (Coulomb) forces relevant to condensed matter and quantum chemistry \cite{paulinet,ferminet, spencer_2020, 2023FrP....1161580N}. The application of these ideas to systems with short-range forces is more recent, an effort driven largely by questions in nuclear physics \cite{saito,lovato1,lovato2,lovato3,lovato4,lovato5,lovato6,Wen:2025mlq,Keeble:2023rre,Keeble:2019bkv,Wang:2024ynn,2023JPCA..127.9159D,2023FrP....1161580N,2023ChPhC..47e4104P,2020JPSJ...89e4706N}. 

In \cite{fermionpaper} some of the present authors considered spin-$\tfrac{1}{2}$ fermions in one-dimensional harmonic traps using neural network ans\"atze. A new network architecture, Kolmogorov-Arnold networks, was used in VMC for the first time in \cite{previous_kan}. The main goal of the present paper is to study a much improved version of the ans\"atze in \cite{fermionpaper} based on Kolmogorov-Arnold networks.

\section{Model and previous results}
We consider spin-$\tfrac{1}{2}$ fermions in a one-dimensional harmonic trap, interacting through a short range potential modeled by a delta function potential between particles of opposite spin. The Hamiltonian for the system is given by:

\beq
H = \sum_{i=1}^{\nup}\sum_{s=\uparrow,\downarrow}
\left( -\cfrac{\hbar^2}{2m} \cfrac{\partial^2}{\partial x_i^{s\ 2}} + \cfrac{m \omega^2}{2}  x_i^{s\ 2} \right) 
+ g
\sum_{i=1}^{\nup}\sum_{j=1}^{\nd}  \delta(\xup_i - \xd_j).
\eeq 
We work in units where $\hbar=m=\omega=1$. The number of up (down) spin particles is $\nup$ ($\nd$) and will consider unpolarized systems with $N_\uparrow=N_\downarrow$ as well as $N_\uparrow \neq N_\downarrow$ systems and for both attractive ($g < 0$) and repulsive cases ($g > 0$). 

Some exact results about this system are known. 
For $g=0$, the ground state energy is given by the sum of the occupied oscillator states: $E_\text{free}=\frac{1}{2}(\nup^2+\nd^2)$.  In the opposite limit,  $g \rightarrow \infty$, the wavefunction vanishes
when particles of the {\it same} spin overlap, just like
the wavefunction of a system where spin up and down particles were identical fermions. For this reason,
the ground state energy is  $E_\infty = \frac{1}{2}(N_\uparrow + N_\downarrow)^2$ \cite{Astra_2016, Guan_2009}. This marks an upper bound for energy of an infinitely repulsive potential, but there appears to be no such analogue in the attractive case. As $g\rightarrow -\infty$, particles of different species pair into tight bosonic bound states (dimers) with binding energy $B_2= mg^2/(4\hbar^2) $ \cite{hans} which dominates the energy of the system. These limiting cases, as well as the result from first-order perturbation theory, are compared to some of our results in \fig{fig:combined}. 
The case $\nup=\nd=1$ is exactly solvable \cite{busch_1998} and  the ``impurity" case, where a single spin-down particle shares the harmonic trap with $N_\uparrow$ fermions, has  an approximate  expression \cite{impurity_dmc} based on the exact  ground state energy of the homogeneous system \cite{mcguire}. 
Finally, perturbative results valid for $|g|\alt 1$ are also easily obtained and will be used as a check below.

Besides  these analytical results, a number of numerical studies explored some regions of the parameter space \cite{Sowi_ski_2019, scottnormalizingflows}. In the strong coupling regime our model is equivalent to a bosonic effective theory \cite{hans} and can also be  mapped  into the Heisenberg spin chain \cite{Levinsen:2015pxc}. 

In particular, some of the present authors have performed a Variational Monte Carlo calculation of this very same system in the unpolarized case ($\nup=\nd$) and in the impurity case ($\nup\geq\nd = 1$) using a wavefunction ans\"atz written in terms of neural networks \cite{fermionpaper}. The present paper is a substantial improvement upon that paper on several aspects.

\section{ans\"atz }

\begin{figure}
    \centering   \includegraphics[width=0.6\linewidth]{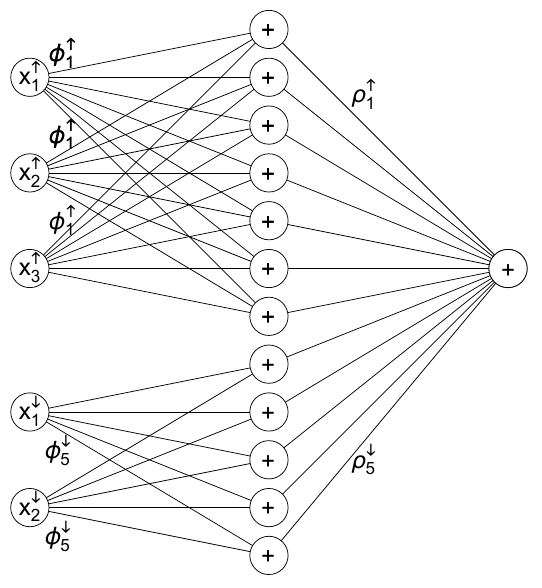}
    \caption{Diagram depicting the Kolmogorov-Arnold neural network representing the function $\kappa(\Xup, \Xd)$ in the $\nup=3, \nd=2$ case. Only some of the lines representing the functions $\phi_{ij}^{\uparrow\downarrow}, \rho_i$ are labelled.} 
    \label{fig:kan}
\end{figure}

The construction of our ans\"atz obeys two principles. The first is {\it universality}, that is,  our ans\"atze should be able to represent any allowed wavefunction as the number of network parameters is increased. This guarantees the asymptotic {\it exactness} of the method. The second is that it should easily represent wavefunctions likely to represent the ground state, which helps with the efficiency of the method.
The necessary conditions restricting the class of acceptable ground state wavefunctions are 1) the wavefunction should be antisymmetric under particle exchange of identical particles and 2) it should be real (as a ground state of a time-reversal invariant Hamiltonian). In one dimension, the antisymmetry is simple to implement as it just imposes the the wavefunction must vanish when two identical particles overlap\footnote{The configuration space of $N$ identical fermions in $d$ dimensions form a $Nd$-dimensional space. The condition $\psi=0$ is one condition so, generically, the manifold where the wavefunction vanishes is $Nd-1$-dimensional. On the other hand, the condition that two particles overlap implies that $d$ conditions to be satisfied and the dimension of the manifold where this occurs is $Nd-d$. When $d=1$, an only in that case, the dimension of these manifolds agree and the wavefunction is required to vanish only when identical particles overlap.}. This can be imposed by having two Slater determinants, one for each spin, multiplied by a Jastrow factor symmetric among the the coordinates of same spin particles:
\beq\label{eq:ansatz1}
\Psi(\Xup, \Xd)=
{\text{det}}
\begin{pmatrix}
\chi_1(\xup_1) & \cdots & \chi_1(\xup_\nup) \\
 &  \vdots  &   \\
 \chi_\nup(\xup_1) & \cdots & \chi_\nup(\xup_\nup)
\end{pmatrix}
{\text{det}}
\begin{pmatrix}
\chi_1(\xd_1) & \cdots & \chi_1(\xd_\nd) \\
 &  \vdots  &   \\
 \chi_\nd(\xd_1) & \cdots & \chi_\nd(\xd_\nd)
\end{pmatrix}
 e^{\kappa(\Xup, \Xd)}
\eeq where $\Xup=(\xup_1, \cdots, \xup_\nup)$,  $\Xd=(\xd_1, \cdots, \xd_\nd)$ and $\chi_i(x^s_j)$ is the $i$-th eigenstate of the one-dimensional harmonic oscillator. We choose to parametrize the Jastrow factor using a 
Kolomogorov-Arnold Network (KAN) \cite{MIT-KAN}, based on the Kolmogorov-Arnold representation theorem \cite{kolmogorov, arnold}, as used in \cite{previous_kan}:
\beq\label{eq:kan}
\kappa(\Xup, \Xd) = \sum_{i=1}^{2N+1}\rho_i\left(\tanh\left(\frac{1}{\nup+\nd} 
\sum_{s=\uparrow \downarrow} \sum_j^{N^s} \phi_i^s(\tanh(x^s_j / a))\right)\right),
\eeq where $\rho_i, \phi^{\uparrow,\downarrow}$
are functions of a single variable. Notice that, according to the Kolmogorov-Arnold theorem, a {\it finite} number of functions $\rho_i, \phi^s_i$ is enough to achieve universality. We parametrize them by quadratic splines with $K$ knots each and defined in the $(-1,1)$ domain. The $\tanh(\dots)$ is included to guarantee that their arguments lie in their domain. A quadratic spline with $K$ ``knots'' is a piecewise function with equally spaced $K$ points from $(-1,1)$ forming $K-1$ intervals. Each interval is a general quadratic polynomial. We require the entire function to be continuous and continuously differentiable which sets some constraints on the coefficients on each interval. It suffices to specify the $y$ coordinates of the function on each of the $K$ knot points to uniquely identify the spline, so each spline has $K$ parameters.
A graphical representation of the KAN in \eq{eq:kan} is shown in \fig{fig:kan} where the single-variable functions are represented by lines while the summation is represented by circles \footnote{This is quite different from the usual graphic representation of feed-forward/multi-layer perceptron networks where functions are represented by circles.}.


 The structure of \eq{eq:kan} enforces the symmetry of $\kappa(\Xup, \Xd)$ under the exchange of the spin up coordinates and the spin down coordinates separately. Except for this restriction, the Kolmogorov-Arnold theorem guarantees that $\kappa(\Xup, \Xd)$ is completely general. Consequently, $e^{\kappa(\Xup, \Xd)}$ is a generic positive function, as required.
It has been debated whether the application of the Kolmogorov-Arnold theorem to represent multivariate functions in terms of single-variable 
requires complicated fractal-like structures in the single-variable functions which poses problems in using them for machine learning tasks \cite{kurkova_1991,girosi_1989}. It is unknown to us whether imposing further regularity constraints on $\kappa(\Xup, \Xd)$ also constrains the line-functions to be smoother. Our previous work with KANs \cite{previous_kan}, the demonstrations in \cite{MIT-KAN, otherkan}, and the results of the present paper do suggest that this is the case.

While the ans\"atz defined by \eq{eq:ansatz1} and \eq{eq:kan} is universal and will produce the ground state in a variational calculation (with large enough $K$), it is not very efficient. The problem arises because the $\delta$-function interaction generates a cusp in the wavefunction of the form:
\beq\label{eq:cusp}
\Psi(\Xup, \Xd) 
\underset{\xup_i \approx \xd_j}{\approx}
\frac{g}{2} \left|\xup_i-\xd_j\right| 
\eeq that is difficult (but not impossible) to reproduce with this ans\"atz. This is the $\delta$-function potential analogue of the Kato condition \cite{kato} in Coulombic systems. For that reason, we modified the original ans\"atz to explicitly include the cusp behavior in the ans\"atz, therefore allowing the function $\kappa(\Xup, \Xd)$ to be smoother and more easily trained. Our final ans\"atz is then:
\beq\label{eq:ansatz2}
\Psi(\Xup, \Xd)=
{\text{det}}
\begin{pmatrix}
\chi_1(\xup_1) & \cdots & \chi_1(\xup_\nup) \\
 &  \vdots  &   \\
 \chi_\nup(\xup_1) & \cdots & \chi_\nup(\xup_\nup)
\end{pmatrix}
{\text{det}}
\begin{pmatrix}
\chi_1(\xd_1) & \cdots & \chi_1(\xd_\nd) \\
 &  \vdots  &   \\
 \chi_\nd(\xd_1) & \cdots & \chi_\nd(\xd_\nd)
\end{pmatrix}
 e^{\kappa(\Xup, \Xd)+\Omega(\Xup, \Xd)}
\eeq where $\kappa(\Xup, \Xd)$ is given by \eq{eq:kan} and
\beq\label{eq:omega}
\Omega(\Xup, \Xd) = \frac{g}{2}
\sum_{i=1}^{\nup} \sum_{j=1}^{\nd} 
|\xup_i-\xd_j| \ 
\text{exp}\left(-\left( \frac{\xup_i-\xd_j}{r_c} \right)^2\right)
\eeq where $r_c$ is another variational parameter. Notice that if the correct ground state wavefunction does not have the behavior shown in \eq{eq:cusp}, the minimization process will  erase the explicit cusp at the cost of a much longer minimization process. This is not what happened in our calculations.

\noindent We estimate the energy by performing a Monte Carlo evaluation of the integral:
\beq\label{eq:energy}
\braket{E} = \frac{\int d\Xup d\Xd\  |\Psi(\Xup, \Xd)|^2 \frac{1}{\Psi(\Xup, \Xd)} H\Psi(\Xup, \Xd)}
{\int d\Xup d\Xd\  |\Psi(\Xup, \Xd)|^2 }
=
\left\langle \frac{1}{\Psi}H\Psi \right\rangle,
\eeq where the bracket $\braket{\cdots}$ denotes the average in relation to the probability measure $\sim |\Psi|^2$. As the determinants and exponentials in the ans\"atz become very large, it is more numerically stable to rewrite \eq{eq:energy} in terms of $\ln \Psi$:
\beq
\braket{E} = \left\langle-\cfrac{1}{2}  (\nabla\log\Psi)^2 
-\cfrac{1}{2}\nabla^2\log\Psi  + \cfrac{1}{2}
\sum_{i=1}^\nup x^{\uparrow 2}_i 
+ \cfrac{1}{2}
\sum_{j=1}^\nd x_j^{\downarrow 2} +g\sum_{i=1}^\nup \sum_{j=1}^\nd \delta(x^{\uparrow}_i-x^{\downarrow}_j)\right\rangle
\eeq
where $\nabla$ denotes spatial derivative in relation to all $\nup+\nd$ coordinates. The Laplacian term in the kinetic energy has a $\delta$-function part coming from the $|\xup_i-\xd_j|$ term which cancels the interaction term exactly\cite{kato}:
\beq
-\frac{1}{2}\braket{\nabla^2 \Omega(\xup,\xd)} = -g \sum_{i,j} \delta(x_i^{\uparrow} - x_j^{\downarrow}) + \text{non-singular terms}
\eeq In fact, that was the original motivation to include the $\Omega(\xup, \xd)$ term in the ans\"atz. We finally obtain:
\beq\label{eq:energy2}
\braket{E} = \left\langle -\cfrac{1}{2} \left( \left(\nabla\log\Psi\right)^2 + (\nabla^2\log\Psi)_{\text{non-sing.}}\right) + \cfrac{1}{2}
\sum_{i=1}^\nup x^{\uparrow 2}_i
+ \cfrac{1}{2}
\sum_{j=1}^\nd x^{\downarrow 2}_j
\right\rangle
\eeq The cancellation above brings another advantage in addition to simplicity. If the  $\delta$-function potential  appeared in \eq{eq:ansatz2}, only a minute fraction of the Monte Carlo samples would probe it, leading to a severe overlap problem. Since it does not appears in \eq{eq:ansatz2}, this problem is bypassed. Another method to deal with the overlap problem caused by strong short-ranged potentials is discussed in \cite{neuralnetworkpaper}.

\section{Minimization and Numerical Results}

We use the estimator in \eq{eq:ansatz2} for computing $\braket{\frac{\partial E}{\partial\theta}}$, where $\theta$ are the variational parameters of the ans\"atz, namely, the values of the splines defining the functions $\phi^{\uparrow\downarrow}_i, \rho_i$  at the knot points, $r_c$ (in \eq{eq:omega}) and $a$ (in \eq{eq:kan}). Several gradient-based algorithms for the minimization of functions of many variables were developed recently, mostly motivated by machine learning tasks. We chose the ADAM algorithm \cite{adam} with learning rates ranging from $10^{-3}$ to $10^{-5}$.

We initialize the ans\"atz with  random parameters for each spline chosen  from the uniform distribution in the range $(-1,1)$;  $r_c=1$ and $a=1$. 
As is typical with machine learning approaches, the specific design of the training procedure is important for the correct and efficient training. 
The Metropolis chain at each training step is started with the last configuration sampled in the previous training step. Since the training steps are small, the wavefunction changes little from one step to the next and the last configuration in one training step is close to be thermalized according to the probability distribution of the next step and only about 1000 Metropolis steps are used for thermalization. 
As the minimization process can afford to be rougher in the early stages of training, the number of Metropolis samples starts small and  increases as the training progresses. About $\approx 10000$ samples is a typical number during training but the final measurement of the energy is done with $64000$ samples. The  noise in the gradient during training is known not to be a decisive factor and may, in fact,  help the algorithm to avoid local minima. In any case, multiple local minima are not expected in the parameter landscape. In fact, consider the energy function in the infinite dimensional space of all wavefunctions. The ground state is unique and the points in parameter space corresponding to excited states have at least one direction along which the energy decreases. The only way that local minima can arise is if the manifold reachable by the ans\"atz with a finite number of parameters is perpendicular to these unstable directions. As the number of parameters increases this scenario becomes less and less likely. We found no evidence of the minimization algorithm being trapped in a local minimum in our calculations. The training proceeds until the energy stabilizes within a $1\%$ uncertainty, an arbitrary number that we therefore claim to be our precision (in some cases we trained to a precision of $\approx 0.1\%$). 


\begin{figure}
    \centering   \includegraphics[width=0.6\linewidth]{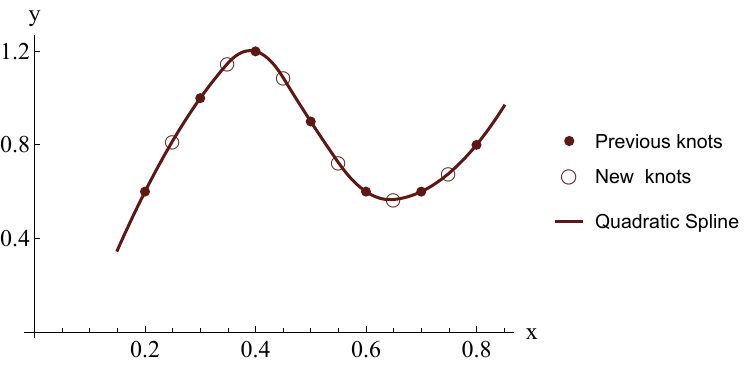}
    \caption{Quadratic spline specified by the values of $y$ at the knots (solid circles) and the new knots introduced during training.} 
    \label{fig:spline}
\end{figure}


We then double the number of spline knots $K$, placing the new knots in the middle of the intervals defined by the previous knots and setting the value of the spline at the new knots to be initially equal to the value of the coarser spline at those points (see \fig{fig:spline}). 
Further training steps are then performed
to verify that no further energy reduction is possible. Typical values of $K=160$ are more than enough to achieve the precision level we aim at for the calculations discussed in this paper. 
When computing the energy for several values of the coupling $g$, it is possible to speed up the process somewhat by  starting the training for one particular value of $g$ with the variational parameters corresponding to the optimal values at a similar value of $g$.  Contrary to the ans\"atz in \cite{fermionpaper} this form of transfer learning does not increase the efficiency substantially and is largely optional. This method of training is inspired from our previous work in \cite{previous_kan} and similar to methods in \cite{saito, fermionpaper, neuralnetworkpaper}. These calculations do not require extensive computational resources. The $\nup=\nd=15$ case, for instance,  can be done in a few hours on a workstation with a 64-core AMD3970X CPU.


\begin{figure}
    \centering   \includegraphics[width=0.6\linewidth]{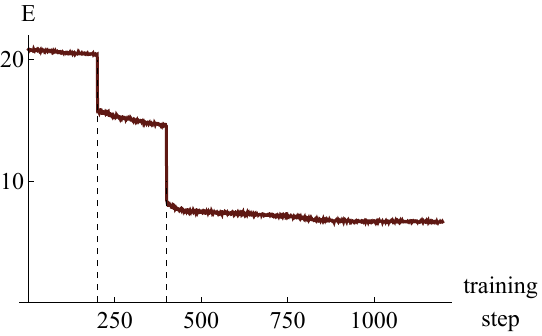}
    \caption{Energy as a function of training step for the $\nup=\nd=5$ case. The coupling and the number of splines changes during the training at the steps shown by the dashed lines: $g=-1, K=10$ in steps $0$ to $200$, 
    $g=-2, K=20$ in steps $200$ to $400$ and
    $g=-3, K=40$ in steps $400$ to $1200$.  Error bars represent only the statistical noise and are not visible on this scale.  
    }
    \label{fig:training}
\end{figure}

A typical training process is depicted on \fig{fig:training}. It must be stressed that in order to make the whole process efficient, some care must be taken in coordinating the increase of the number of knots $K$ and the number of Monte Carlo samples.  As the precision provided by larger $K$ is possible, a concomitant increase in the number of Monte Carlo steps is required.
In all our calculations we have not encountered a single wavefunction that seemed to be a {\it local} minimum of the energy.

\section{Results and Discussion}

\begin{table}
    \centering
    \begin{minipage}{0.32\textwidth}
        \centering
        \begin{tabular}{|c|c|c|}
        \hline
        \multicolumn{3}{|c|}{$\nup=\nd=1$} \\
        \hline
        $g$ & this paper & exact \\ 
        \hline
        -3 & -1.694(8) & -1.698 \\    
        \hline
        -2 & -0.395(7) & -0.399 \\   
        \hline
        -1 & 0.471(4) & 0.470 \\ 
        \hline
        1 & 1.308(5) & 1.307 \\   
        \hline
        2 & 1.497(12) & 1.487 \\   
        \hline
        3 & 1.605(10) & 1.600 \\   
        \hline
        \end{tabular}
        \label{tab:busch1}
    \end{minipage}
    \hfill
    \begin{minipage}{0.32\textwidth}
        \centering
        \begin{tabular}{|c|c|c|}
        \hline
        \multicolumn{3}{|c|}{$\nup=\nd=2$} \\
        \hline
        $g$ & this paper & exact \\ 
        \hline
        0.5 & 4.509(5) & 4.503 \\    
        \hline
        -0.5 & 3.402(4) & 3.405 \\   
        \hline
        \end{tabular}
        \label{tab:busch2}
    \end{minipage}
    \hfill
    \begin{minipage}{0.32\textwidth}
        \centering
        \begin{tabular}{|c|c|c|}
        \hline
        \multicolumn{3}{|c|}{$\nup=\nd=3$} \\
        \hline
        $g$ & this paper & exact \\ 
        \hline
        0.5 & 9.940(6) & 9.938 \\    
        \hline
        -0.5 & 7.934(7) & 7.931 \\   
        \hline
        \end{tabular}
        \label{tab:busch3}
    \end{minipage}
    \caption{Comparison of energies to analytic results from \cite{busch_1998} for different $\nup=\nd$ values. In these small systems the training process was pushed to guarantee a precision of $0.01\%$ in order to stress test the method. The quoted uncertainties reflect statistical noise only.}
    \label{tab:busch}
\end{table}

Using the ans\"atz and training method discussed above we performed several calculation with different numbers of particles and values of the coupling $g$. 
We stop training when upon increasing the number of parameters and the statistics in the Monte Carlo, the energy decreases by less than 1\% which becomes our estimate of systematic error. In some cases, for testing purposes, we push the precision to $0.1\%$.  

\begin{figure}[t]
    \centering
    \includegraphics[width=0.45\linewidth]{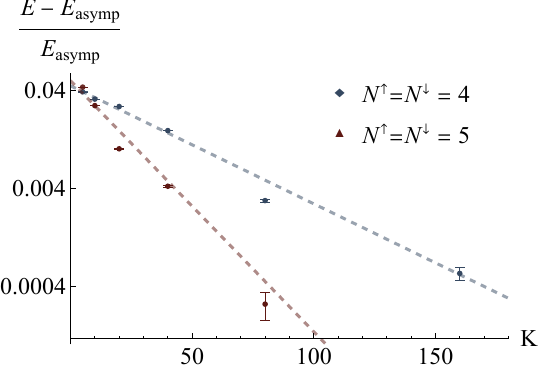}
    \hfill
    \includegraphics[width=0.45\linewidth]{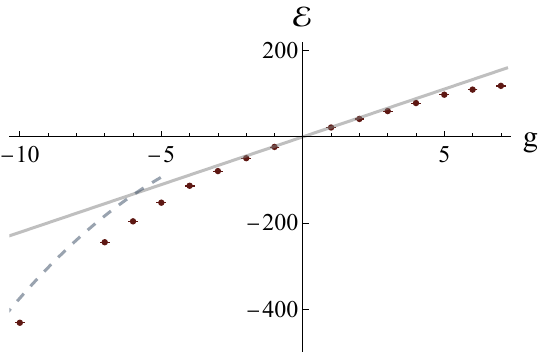}
    \caption{Left: Log plot of the relative error in the ground state energy of the $g=3.0$ case, $N_\uparrow=N_\downarrow=5$, (in red) and $\nup=\nd=4$ (in blue) system. We take $E_{\text{asym}}$ as the value obtained at $K=320$ ($K=160$). The fits, shown as dashed lines, suggest an exponential increase of precision with the error scaling as $\sim e^{-K/35.9} (e^{-K/16.9})$. Right: Interaction energy $\mathcal{E}$ for $N_\uparrow=N_\downarrow=15$ particles plotted against the coupling strength $g$. The gray line is the result from first order perturbation theory and the dashed line represents the dimer limit. The error bars include the statistical uncertainty only.}
    \label{fig:combined}
\end{figure}

\begin{figure}[t]     \includegraphics[width=0.6\linewidth]{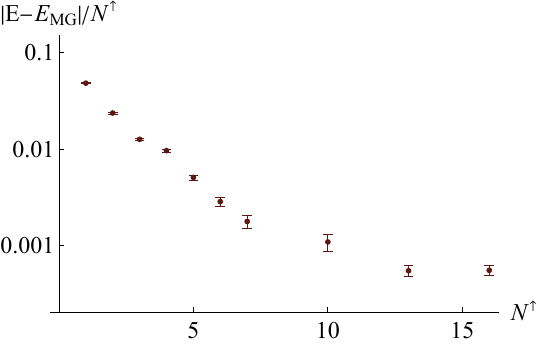}
    \caption{Log plot of the difference between the energy per particle given by our results and the McGuire formula \cite{mcguire}, as a function of $\nup$, with $\nd=1$, $g=3$. }   
    \label{fig:impurity}
\end{figure}

First, we validated our methods by comparing  a variety of known results to ours. 
An analytic solution to the $N_\uparrow=N_\downarrow=1$ case is shown in \cite{busch_1998}. We compare it in the left panel in  \tab{tab:busch}. We can also obtain exact results for $\nup= \nd=2,3$ by direct diagonalization of the Hamiltonian in the oscillator basis (with oscillator level $n$ truncated to some large value $n\leq \Lambda$.). These results are also shown in the central and right panels of \tab{tab:busch}. 
In all these comparisons with small system sizes, we pushed the training further than usual  in order to stress test our approach. We train for 100 steps beyond the point where the energy does not change by more than $\approx 0.1\%$, which becomes our estimate of systematic error in the procedure. We find that, in all these cases, we agree with the established results within the estimated precision.
A further check can be made in the ``impurity" case, where $\nd=1$ and $\nup$ varies. As previously mentioned, a phenomenological analytic expression \cite{impurity_dmc} based on the exact result in a homogeneous system, the McGuire formula \cite{mcguire}, is known to reproduce the ground state energy very accurately, especially at large system sizes. The result of this comparison is shown in \fig{fig:impurity} which shows our results approaching the McGuire formula quickly as $\nup$ is increased. A further comparison with perturbative results will be  shown later in \fig{fig:combined}.


\begin{figure}[t]
    \centering
    \begin{minipage}{0.48\textwidth}
        \centering
        \includegraphics[width=\linewidth]{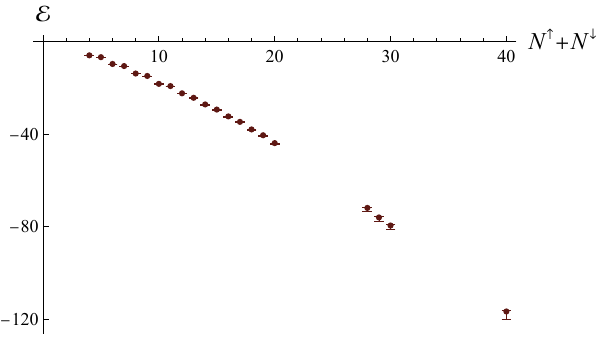}
    \end{minipage}
    \hfill
    \begin{minipage}{0.48\textwidth}
        \centering
        \includegraphics[width=\linewidth]{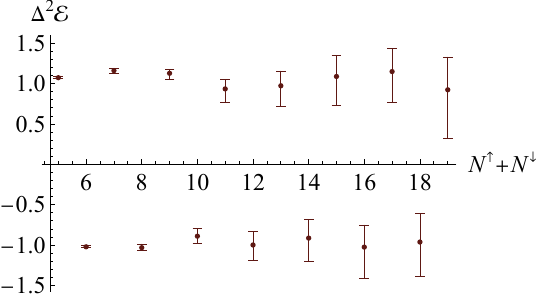}
    \end{minipage}
    \caption{{\it left panel:} Interaction energy per particle of the ground state $\mathcal{E}=E-E_{\text{osc}}$ as a function of the total number of particles $\nup+\nd$. {\it right panel:}  $\Delta^2 \mathcal{E} = \mathcal{E}(\nup, \nup)-\frac{1}{2}(\mathcal{E}(\nup+1, \nup)+\mathcal{E}(\nup-1, \nup))$, as a function of $\nup$ for $g=-3.0$. The even-odd pairing effect is evident. The error bars include only the statistical uncertainties and a one-sided $1\%$ theoretical error (since the variational bound guarantees a rigorous upper bound in the energy) at each value of the energy. Notice that the $\approx 1\%$ uncertainty is enhanced by the particular combination contributing to $\Delta^2\mathcal{E}$, especially at larger $\nup$.}   
    \label{fig:pairing_combined}
\end{figure}

For attractive interactions ($g<0$) the most distinctive feature is the pairing between spin up and spin down particles. This is evidenced in \fig{fig:pairing_combined} where we plot  a common measure of gap effects,  the combination:
\beq\label{eq:gap}
\Delta^2 \mathcal{E} = \mathcal{E}(\nup, \nup)-\cfrac{1}{2}(\mathcal{E}(\nup+1, \nup)+\mathcal{E}(\nup-1, \nup)),
\eeq where $\mathcal{E}$ is the {\it interaction} energy (not including the oscillator part).  The combination $\Delta^2 \mathcal{E}$ distinguishes the nearby even-odd ones (and vice-versa). The gap is expected to approach a constant in the $\nup=\nd\rightarrow\infty$ limit while the individual energies for fixed $\nup, \nd$ grow with system size. Therefore, a certain degradation of precision results from the cancellation among the terms in \eq{eq:gap}. Our results strongly suggest a gap independent of system size as one would expect for a system with constant density of states \cite{Sowi_ski_2019, Grining_2015_gap}.


We consider now larger systems, where the use of KAN-based ans\"atze can be advantageous.
In the right panel of \fig{fig:combined} we show our results for the $\nup=\nd=15$ case as a function of the coupling $g$. At small values of the coupling $|g|\leq 1$ there is agreement with the first-order perturbation theory result (gray line), what can be viewed as a further validation of our calculations. At large negative values of $g$, when spin up and down particles form tightly bound bosonic dimers with binding energy $B_2=g^2/(4\hbar^2)$, our results slowly approach the  limit $E \approx -\nup g^2/4$ (the dashed line in the right panel of \fig{fig:combined}), as expected.

While our ans\"atz can, in principle, represent any allowed wavefunction, the fact that we use spline with a finite number $K$ of knots limits its representability. A natural question then is how close we can get to the ground state for a given number of knots and, perhaps more importantly, how the precision scales with $K$. In order to answer this question we study a relatively small system ($\nup=\nd=5$) at moderate coupling ($g=3.0$) and larger precision ($\approx 0.1\%$) as a function of the number of knots. For each value of $K$ we used a large enough number of Monte Carlo samples and train it for long enough so that any imprecision can be assigned exclusively to the lack of representability of the ans\"atz. The result is shown in the left panel of \fig{fig:combined} where the relative error of the ground state energy is plotted against the number of knots (we took the ``exact" energy as the one obtained with $K=320$ which differ from the $K=160$ by $0.04\%$ only, below the statistical error.). The data suggests, at least in these cases, an exponential increase in precision with $K$.


\section{Conclusion}
We studied a neural network-based variational ans\"atz  for a one-dimensional system of spin-$\tfrac{1}{2}$ particles in a harmonic trap. The choice of ans\"atz  is based on two principles: universality (as the number of parameters increases) and the known short-distance properties of the wavefunction. The first guarantees the exactness of the method while the second contributes to its efficiency. The way universality can be guaranteed relies heavily on technology coming from the machine learning/artificial intelligence field. In particular, we use a new kind of neural architecture (Kolmogorov-Arnold networks) that offers several advantages in our problem.

The incorporation of the short-distance asymptotics in the wavefunction while maintaining universality is one of the advancements in the method described here. It is likely to be useful in other contexts where strong, short-range potentials are important, such as nuclear physics. For finite potentials, the asymptotic behavior of the wavefunction, contrary to the case considered here, depends on the energy $E$ but only weakly and one can envisage an expansion in powers of $E$, in the spirit of effective theories \cite{eft-review}. The implementation of this idea, in a three-dimensional model, is the next step of our program.

It is important to understand the scaling of the computational cost of every new numerical technique. In our case, there are three steps of the process that should be separated. The first is the Monte Carlo evaluation of the energies and gradients, {\it given a wavefunction}. This is a well understood step whose scaling is well understood -- the precision scales like the square root of the number of samples taken (uncertainty $\sim 1/\sqrt{n_{samples}}$). One issue that could arise here -- the breakdown of ergodicity-- was not found despite  constant monitoring. The second is the cost of training. Again, we have no evidence of the training ``getting stuck" in a local minimum. We argued that for generic enough ans\"atze we don't expect the existence of local minima. Finally, there is a cost associated with the number of parameters necessary to represent the correct ground state wavefunction. We have some preliminary evidence that the representability, measured by the error in the ground state energy,  grows exponentially with the number of spline knots. If confirmed, this means that the error in the method is dominated by the number of samples used, just like any other ``exact" Monte Carlo method. This elevates VMC to the status of  ``exact" method. Of course, well known difficulties appear in the extension of these ideas to three dimensions, mostly related to the location of the zeros of the wavefunction, are the topic of current research and will be reported elsewhere.

\section{Acknowledgments}
This work was supported in part by the U.S. Department of Energy, Office of Nuclear Physics under
Award Number DE-FG02-93ER40762. S.P. was partially supported by DOE Grant KA2401045. H.K. was partially supported by the DOE, Office of Science, Office of High Energy Physics (HEP), HEP Computing Traineeship program: Lattice Gauge Theory for HEP (under grant no. DE-SC0024053).

\bibliography{1D-trapped-fermions}
\end{document}